\documentclass[sigconf]{acmart}
\usepackage{graphicx} 
\usepackage{subfigure} 
\usepackage{caption}
\usepackage{soul}
\usepackage{color}
\setcopyright{none}
\renewcommand\footnotetextcopyrightpermission[1]{} 
\AtBeginDocument{%
  \providecommand\BibTeX{{%
    \normalfont B\kern-0.5em{\scshape i\kern-0.25em b}\kern-0.8em\TeX}}}

\begin{document}

\title{Verification of $\mathcal{L}_1$ Adaptive Control using Verse Library: A Case Study of Quadrotors}

\author{Lin Song, Yangge Li, Sheng Cheng, Pan Zhao, Sayan Mitra, Naira Hovakimyan}

\affiliation{University of Illinois Urbana-Champaign, United States
  \state{}
  \country{}}
\renewcommand{\shortauthors}{Song and Li, et al.}
\begin{abstract}
  $\mathcal{L}_1$ adaptive control ($\mathcal{L}_1$AC) is a control design technique that can  handle a broad class of system uncertainties and provide transient performance guarantees. 
In this work-in-progress abstract, we discuss how existing formal verification tools can be applied to check the performance of $\mathcal{L}_1$AC systems.
We show that the theoretical transient performance and robustness guarantees 
of an $\mathcal{L}_1$ adaptive controller for an 18-dimensional quadrotor system can be verified using the recently developed Verse reachability analysis tool. We will further consider the performance verification of $\mathcal{L}_1$AC on systems with learning-enabled components.
\end{abstract}
\maketitle

\footnotetext{This work is funded by NASA ULI grant 80NSSC22M0070 and AFOSR. }





\section{Introduction}

Advanced air mobility (AAM) aims to build a reliable and efficient aviation transportation system using highly automated vehicles, e.g., vertical take-off and landing (VTOL) aircraft. A verification and validation (V\&V) framework is a critical element for AAM to ensure  enhanced level of reliability via formal verification of the controller performance. 

$\mathcal{L}_1$ adaptive controller ($\mathcal{L}_1$AC) distinguishes itself by its capability of compensating for a broad class of model uncertainties with fast adaptation while providing transient and steady-state performance guarantees~\cite{hovakimyan2010l1}. $\mathcal{L}_1$ adaptive control has been successfully deployed and validated on NASA's AirStar 5.5\% subscale generic transport aircraft model~\cite{gregory2009l1}, Calspan's Learjet~\cite{ackerman2017evaluation}, and unmanned aerial vehicles~\cite{wu2023L1QuadFull}. However, there is no prior research  on the formal verification of the transient bounds and robust performance of $\mathcal{L}_1$AC. For complex but deterministic control systems, verification tools, like DryVR~\cite{fan2017dryvr} and the recently-developed Verse~\cite{li2023verse} which can handle black-box components, can be  promising.   
In this paper, we explore applying the Verse Library to formally verify   $\mathcal{L}_1$AC's  robustness and transient performance. From the verification results, we observe that $\mathcal{L}_1$AC achieves verifiable robust transient   performance, and is capable of fast adaptation in systems with time-varying uncertainties. Moreover, we verify that $\mathcal{L}_1$AC guarantees a delay margin (bounded away from zero) when control inputs are subject to time delays, and the tracking performance provided by $\mathcal{L}_1$AC  degrades gracefully as the injected input delay increases.

\section{Problem Formulation}
\paragraph{Scenarios}
We consider the performance verification of $\mathcal{L}_1$AC on an 18-dimensional quadrotor system. 
We use a geometric tracking controller as the baseline controller. For comparison, we implement verification on the drone in two scenarios with different controllers --- baseline geometric control with and without $\mathcal{L}_1$AC. 
\paragraph{Plant --- Quadrotor Dynamics}

The equations of motions (EOMs) of a quadrotor~\cite{wu20221,lee2010geometric} are given by
\begin{equation}\label{eq:eom}
    \dot p = v, 
    \dot v = ge_3 - fRe_3/m,
    \dot R = R\Omega^\wedge, 
    \dot \Omega = J^{-1}(M-\Omega \times  J\Omega),
\end{equation}
 where $p, v \in \mathbb{R}^3$ are the position and velocity of the quadrotor's center of mass (COM) in the inertial frame, $g$ is the gravitational acceleration, $m$ is the vehicle mass, $\Omega \in \mathbb{R}^3$ is the angular velocity in the body-fixed frame, $J \in \mathbb{R}^{3 \times 3}$ is the moment of inertia matrix, and $R \in \{ \mathbb{R}^{3 \times 3}|R^\top R = I, \det(R) = 1\}$ is the rotation matrix. The wedge operator $(\cdot)^{\wedge}: \mathbb{R}^3 \to \mathfrak{so}(3)$ denotes the mapping to the space of skew-symmetric matrices. The control inputs include the collective thrust $f $ and the moment $M \in \mathbb{R}^3$ in the body-fixed frame. 


\paragraph{Controller --- Geometric Control with $\mathcal{L}_1$ Augmentation}\label{sec2_ctrl}
The geometric controller~\cite{lee2010geometric} can ensure exponential stability for quadrotor's nominal dynamics while tracking a prescribed trajectory $p_d(t) \in \mathbb{R}^3$ and yaw angle $\psi_d(t) \in \mathbb{R}$ for $t \in [0,t_f]$.  In this paper, we compute the baseline geometric control law for the nominal quadrotor dynamics~\eqref{eq:eom}. The desired thrust is  $f = -F_d \cdot (Re_3)$, where $F_d = -K_pe_p - K_v e_v - mge_3 + m\ddot{p}_d$ denotes the desired force vector, $e_p, e_v$ are the position and velocity error vectors. The desired moment given by the geometric controller is $
     M = -K_R e_R - K_\Omega e_\Omega + \Omega \times J\Omega \nonumber 
     - J(\Omega^\wedge R^\top R_d \Omega_d - R^\top R_d \dot{\Omega}_d)$, where $e_R,e_\Omega, R_d$ are the attitude error, angular velocity error, and desired rotation matrix (see~\cite{wu20221,lee2010geometric} for the computation of $e_R, e_\Omega,R_d,\Omega_d,\dot{\Omega}_d$). 
     To establish verification in the presence of uncertainties, we use the geometric control with $\mathcal{L}_1$ augmentation introduced in~\cite{wu20221} to compensate for the uncertainties. $\mathcal{L}_1$AC includes a state predictor, an adaptation law, and a low-pass filter. (Interested readers can refer to~\cite{wu2023L1QuadFull} for a detailed discussion on the theoretical   bounds guaranteed by the $\mathcal{L}_1$AC.) 
\paragraph{Uncertainty \& Input Delays}
We create a model mismatch setting, i.e., the mass applied for controller design ($m_0$) is different from the actual mass ($m'$) for verification, where $m'$ is also a time-varying uncertain value with known time-dependent bounds. This setting is designed to verify the robust performance of $\mathcal{L}_1$AC and its capability of fast adaptation in systems with time-varying parametric uncertainties. In the use of a verification tool (e.g., Verse~\cite{li2023verse}), we introduce the uncertain mass as an augmented state, such that the time-dependency and uncertain initial masses can be captured in the verification. Time delays on control inputs widely exist on the hardware. In addition to the uncertain model mass, we also inject input delays to verify the robustness margins achieved by $\mathcal{L}_1$AC. 




\section{Verification Solution}

\begin{figure}[h]
  \centering
  \includegraphics[height=2.7cm]{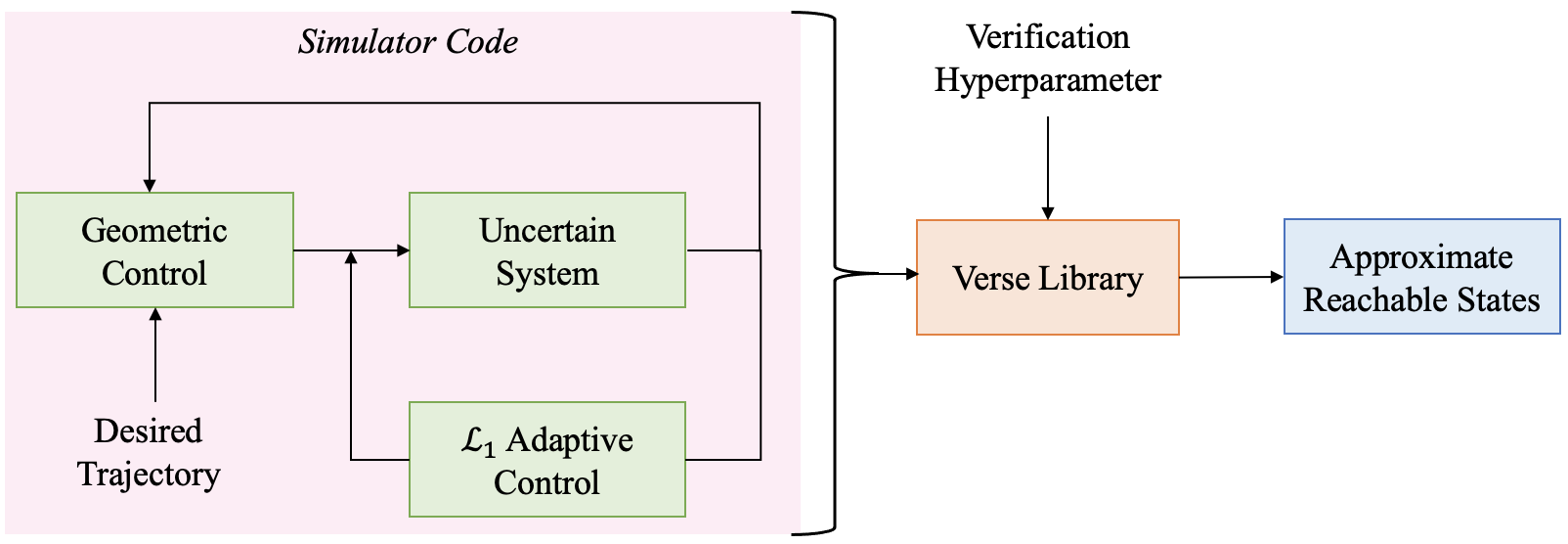}
  \caption{Verification architecture of geometric tracking control with $\mathcal{L}_1$ augmentation using the Verse Library~\cite{li2023verse}.}
  \label{fig:arch}

\end{figure}

\paragraph{Solution of ordinary differential equation (ODE)}
An ordinary differential equation (ODE) describes the instantaneous rule governing the change of states of a physical system.  The solution of the ODE, called trajectory,  models how the system state changes over time. ODE solutions can describe the system states  reachable within a given time horizon, which can also be interpreted from the perspective of trajectory evolution. The complete formulation of the quadrotor's ODE dynamics~\eqref{eq:eom} with geometric controller and $\mathcal{L}_1$AC is omitted, and we refer the interested readers to~\cite{wu2023L1QuadFull} for more details.
\paragraph{Reachability Analysis}
\textcolor{black}{For an ODE system with a set of initial states $X_0$ or parameter values, the {\em reachable set\/}, denoted by $Reach(X_0, t_f)$, is the  set of states that the solutions of the system can hit from {\em any\/} initial state $X_0$ within time $t_f$. 
Given an unsafe set $U$, checking that $Reach(X_0, t_f) \cap U =\emptyset$ is a standard way for checking bounded safety. There are several tools for over-approximating $Reach(X_0, t_f)$ for the above check (see~\cite{ReachSurveyChenNASA22,MitraCPSBook2021} for recent survey).
}


\paragraph{Reachability analysis with Verse}

\textcolor{black}{Verse~\cite{li2023verse} is a Python library for modeling and reachability analysis of hybrid, multi-agent scenarios. It allows for the dynamics of the individual agents to be described by black-box simulators (written in any language), uses the probabilistic algorithm of DryVR~\cite{fan2017dryvr} for computing sensitivity functions with probably approximately correct (PAC) guarantees, and then uses a simulation-based algorithm for over-approximating the reachable states.} 
We present the verification architecture for the quadrotor system with 
 $\mathcal L_1$AC in Fig~\ref{fig:arch}. The simulator code of the closed-loop quadrotor system is taken as the Verse library input and is intepreted as a black-box simulator. The inputs for verification also include some hyperparameters, e.g., time horizon and step size. 


\section{Experiments and Results}

\begin{figure}
\subfigure{\includegraphics[width=4cm]{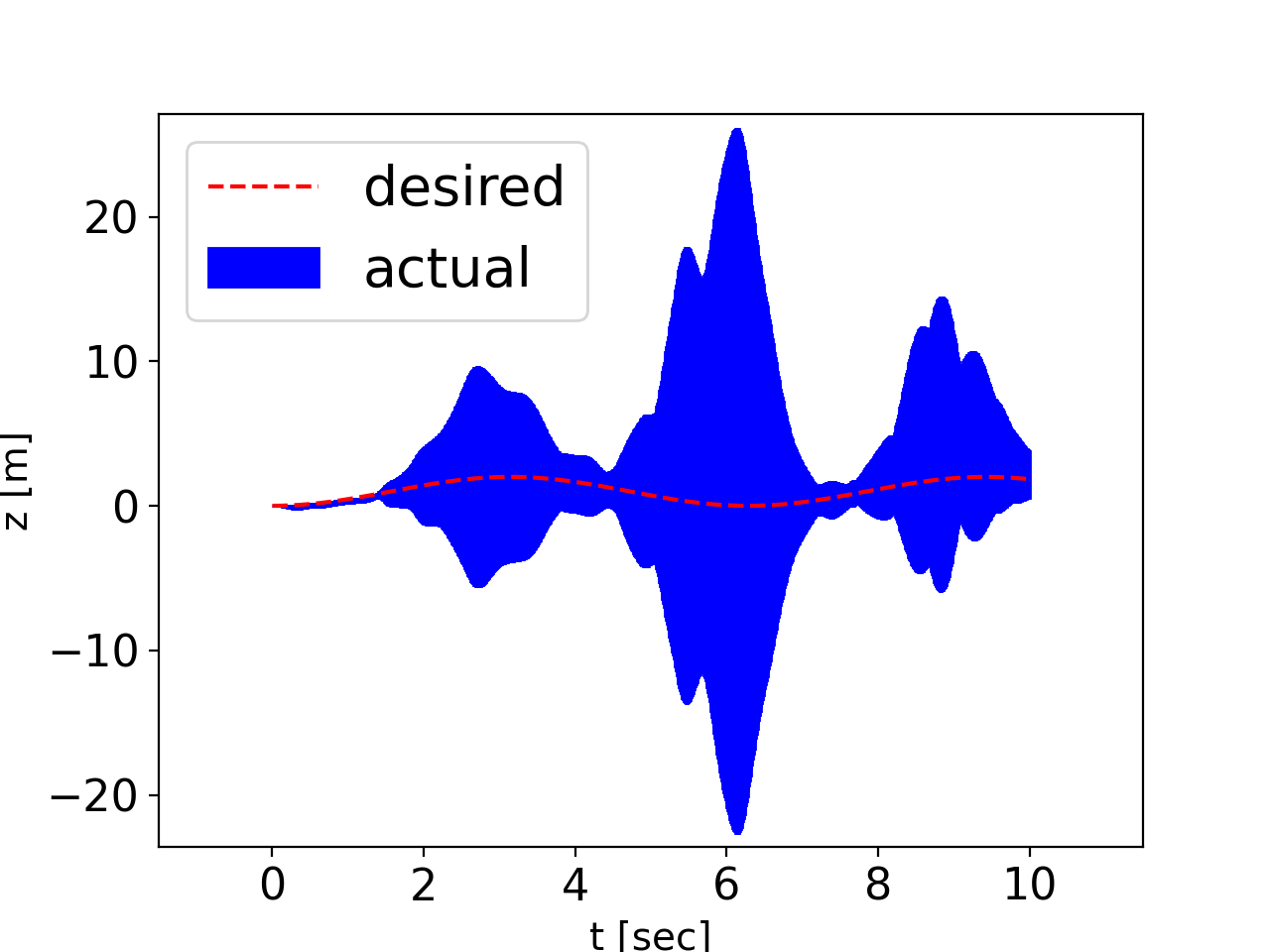}}
\subfigure{\includegraphics[width=4cm]{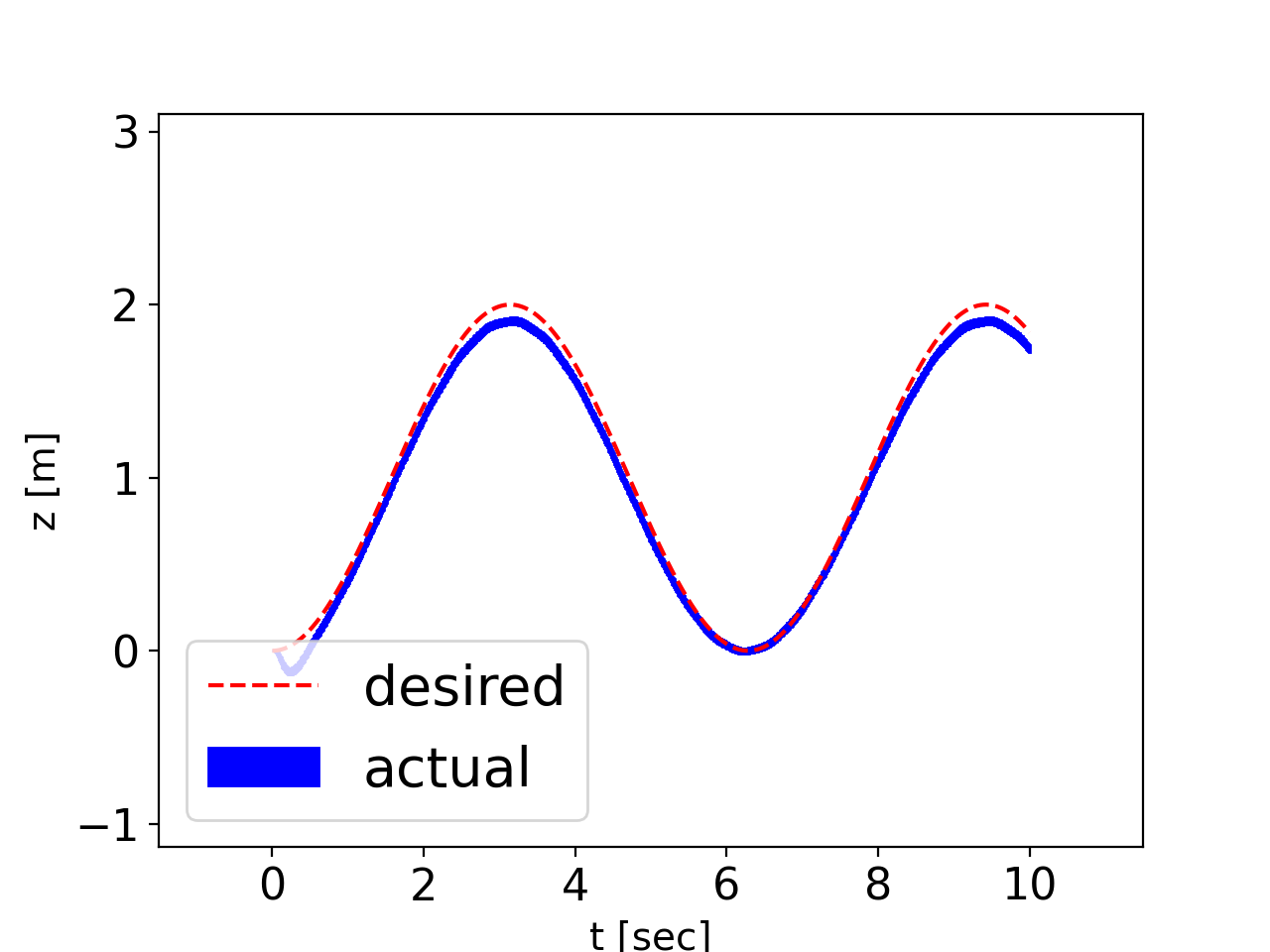}}
\caption{Transient performance verification of $\mathcal{L}_1$AC subject to time-varying system parameters: performance of geometric control (left) and geometric control w/ $\mathcal{L}_1$AC (right)}
\label{fig:fig1_big}
\end{figure}

\begin{figure}
\vspace{-1pc}
\subfigure{\includegraphics[width=4cm]{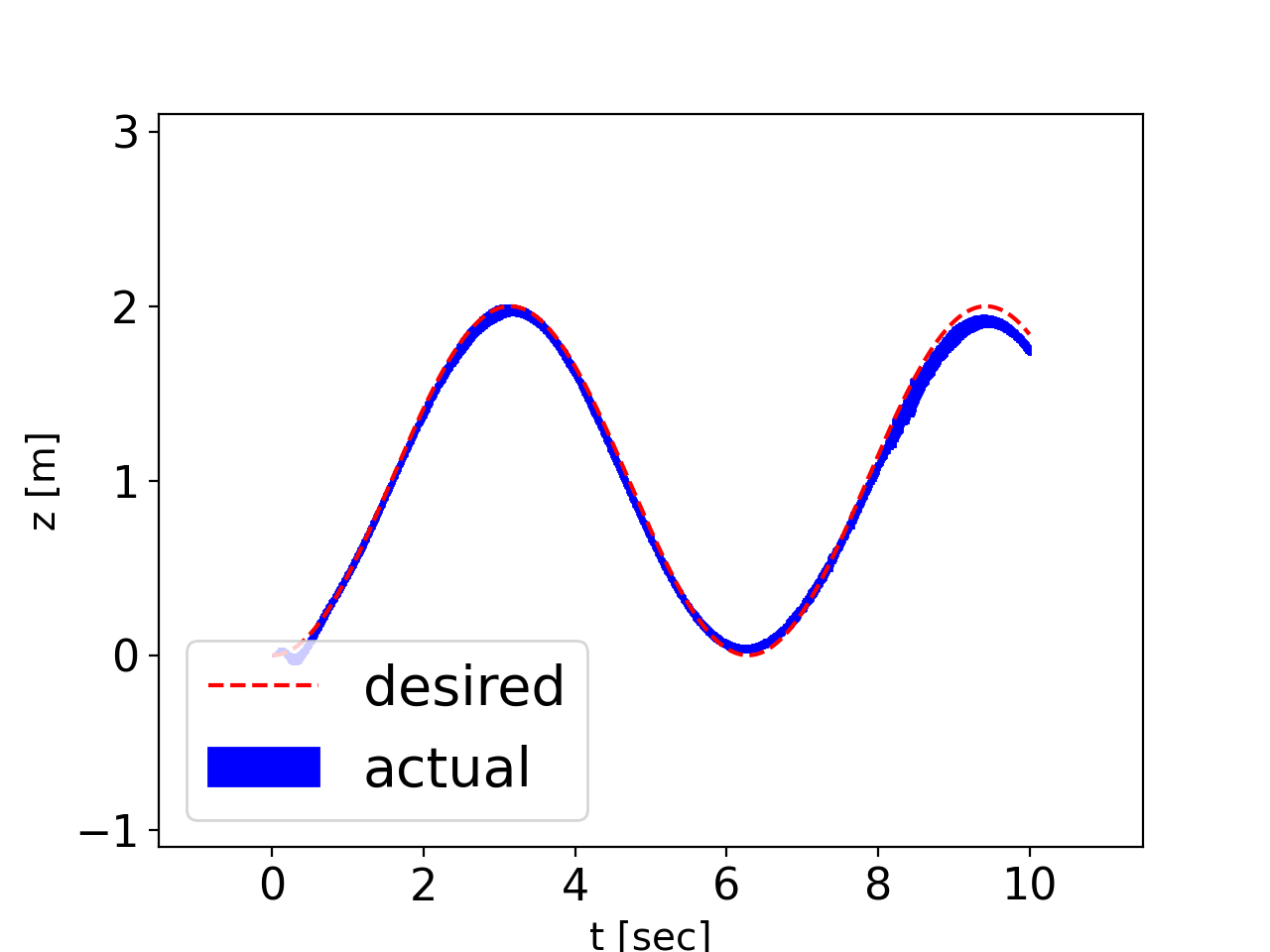}}
\subfigure{\includegraphics[width=4cm]{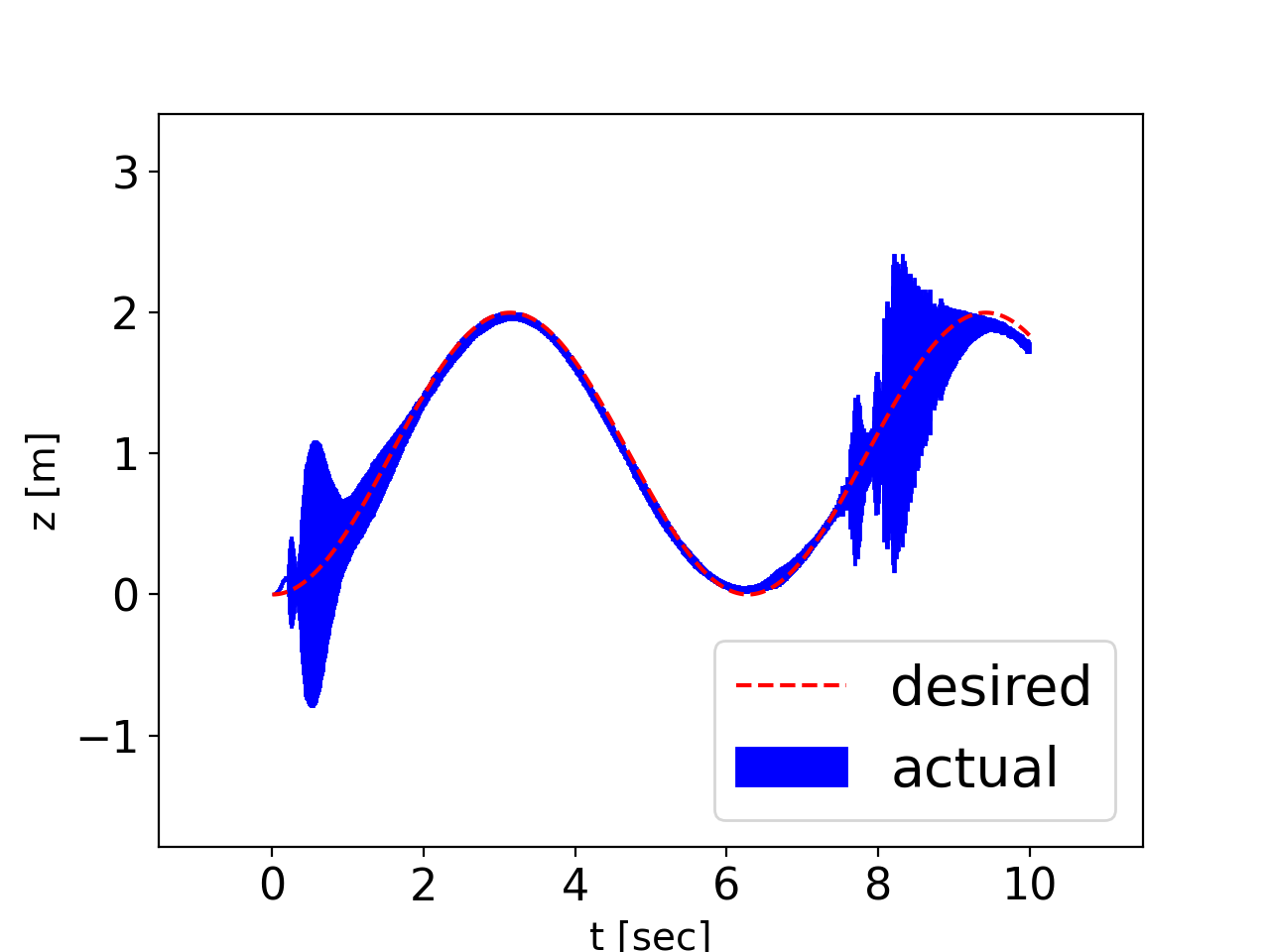}}
\caption{Robust performance verification of $\mathcal{L}_1$AC in the presence of input delay and time-varying system parameters:  with input delay of 60ms (left) and 120ms (right)}
\label{fig:subfigures}
\end{figure}


In the experiments, we evaluate the robust performance of $\mathcal{L}_1$AC by observing the computed reachable states on uncertain systems using Verse. 
 We first verify the transient performance guarantees of $\mathcal{L}_1$AC by modeling the quadrotor mass in~\eqref{eq:eom} as an uncertain and rapidly changing value with known bounds. The control goal for the quadrotor is to track a given reference trajectory. From the computed z-axis reachtube in the cases of with and without $\mathcal{L}_1$AC shown in~Fig.~\ref{fig:fig1_big}, we see that $\mathcal{L}_1$AC achieves consistent and better tracking performance even with time-varying system parameters. $\mathcal{L}_1$AC is capable of handling rapidly changing parametric uncertainties and achieving consistent performance with fast adaptation. Furthermore, we verify the robustness of $\mathcal{L}_1$AC against input delay. We still consider an uncertain time-varying quadrotor mass and gradually increase the injected time delay on the control input. The verification results in the form of \textcolor{black}{reachtube} in the presence of different input delays are shown in Fig.~\ref{fig:subfigures}. 
 It is observed that $\mathcal{L}_1$AC delivers reasonably good performance in the presence of input delay up to 60~ms. The verification results also show that the tracking performance degrades gracefully as the injected time delay increases. 
For future work, we will consider the verification of $\mathcal{L}_1$ adaptive controllers with learning-enabled components.




\bibliographystyle{abbrv}
\bibliography{main-ref}


\end{document}